\documentclass[12pt]{JHEP3}
\usepackage{bm}
\usepackage{epsfig}
\usepackage{citesort}
\usepackage{graphicx}
\usepackage{multirow}
\usepackage{url}
\usepackage{units}
\usepackage{epstopdf}
\usepackage{subfig}

\makeatletter

\title{Suppressing CMB low multipoles with ISW effect}

\author{Santanu Das \& Tarun Souradeep\\
Inter-University Centre for Astronomy and Astrophysics, Post
Bag 4, Ganeshkhind, Pune 411007, India \\
E-mail: \email{santanud@iucaa.ernet.in}, \email{tarun@iucaa.ernet.in}}

\abstract{Recent results of Planck data reveal that the power
 \cite{Ade2013b,Ade2013} in the low multipoles of 
the CMB angular power spectrum, approximately up to $l=30$, is significantly lower than 
the theoretically predicted in the best fit $\Lambda$CDM model. 
There are different known physical effects
that can affect the power at low multipoles, such as features in the primordial power 
spectrum (PPS) in some models of inflation 
and ISW effect. In this paper we investigate the possibility of invoking the Integrated 
Sachs-Wolfe (ISW) effect to explain the 
power deficit at low multipoles. The ISW effect that 
originates from the late time expansion history of the 
universe is rich in possibilities given the limited understanding of the origin of 
dark energy (DE). It is a common understanding 
that the ISW effect adds to the power at the low multipoles of the CMB angular power spectrum. In this 
paper we carry out an analytic study to show that there are some expansion histories 
in which the ISW effect, instead of adding power, provides negative 
contribution to the power at low multipoles. 
Guided by the analytic study, we present examples of  
the features required in the late time expansion history of the 
universe that could explain the power deficiency through the ISW effect. 
We also show that an ISW origin of power deficiency is consistent, at present, with other 
cosmological observations that probe 
the expansion history such as distance modulus, matter power spectrum and the evolution of cluster 
number count. We also show that the ISW effect may be distinguished from 
power deficit originating from features in the PPS using the measurements of the CMB polarization
spectrum at low multipoles expected from Planck. We conclude that the power at low multipoles
of the CMB anisotropy could well be closely linked to Dark Energy puzzle in cosmology and this 
observation could be actually pointing to richer phenomenology of DE beyond the cosmological
constant $\Lambda$. 
}

\begin{document}
The cosmic microwave background (CMB) radiation is one of the most important discovery in the 
field of astronomy. The precision in the measurement of the CMB has improved dramatically 
in last few years. The standard model of the cosmology, can explain almost all the features 
of the CMB power spectrum using a handful set of parameters. This has led to the standard 
cosmological model being well accepted. The power 
spectrum of CMB consists of the contribution from different aspects of the physics in the early universe. 
However, an important contribution to the CMB power spectrum linked to the expansion history of the universe 
after the surface of last scattering arises from the late time ISW effect. The CMB photons from 
the last scattering surface fall in and climb 
out of several gravitational potential wells along their path to the present. But if there is 
no large scale evolution of the potential in the 
universe over time such as in case of the SCDM model, then the net energy of photons will not 
deviate from the mean redshift due to Hubble after the last 
scattering surface to the present era. But in most cosmological models there is evolution
in the amplitude of the potential over time and the energy of 
photon has additional changes after LSS during the free propagation. This effect, called the 
ISW effect\cite{Sachs1967} is a source of CMB temperature anisotropy that depends on the evolution of the gravitational 
potential due to the expansion history of the universe.

The ISW effect has been analyzed by various authors \cite{Sachs1967,Pogosian2005,Corasaniti2003,Cai2008}. An analytical calculation of the ISW 
effect based on the scalar field dark energy model is shown in \cite{Bertacca2007}. Approximate 
analytical expressions for the ISW effect are discussed in \cite{Hu1995,Czinner2005tb}. It is 
commonly believed that the ISW term always increases the power of the CMB power spectrum at 
the low multipoles. But this presumption is not correct when we take into account the effect of 
the cross coupling (``interference'') of the ISW source term with the primordial source term\cite{Kofman1985}. In general for $\Lambda$CDM 
model, the interference term between the ISW source function and the primordial source function is 
negative but very small. %But it is possible to have 
There exist expansion histories of the universe where the cross term of the ISW and primordial source term 
can be large and hence the ISW effect may provide a negative contribution to the low CMB multipoles.

The ISW effect mainly affects the low multipoles of the CMB power spectrum. The recent data release 
of the Planck satellite \cite{Ade2013,Ade2013b} has shown that the integrated power at the low multipoles of 
observed $C_{l}^{TT}$ (mainly for $l<30$) is significantly lower than the theoretical predictions of 
the $\Lambda$CDM model. The origin of this 2.5-3$\sigma$ power deficiency at the low multipoles is not 
satisfactorily understood. Such power deficiency at the low multipole may be
 explained by modifying the inflationary power spectrum \cite{Jain2008,Sinha2006,Contaldi2003,Mortonson2009}. %$$$$$$$$$$$$$$$$$$$$$$$
But that the ISW effect can decrease the 
power of the low multipole is a less known fact\cite{Kofman1985}. In this paper our goal is to explore the 
theoretical possibility that for some expansion history ISW can provide the 
negative contribution to the power at the low CMB multipoles and to find out the features required 
in the expansion history of the universe that can provide this effect. In this paper, using a new line of 
sight code CMBAns \cite{Das-2010,Das2012} we explore the features required 
in the expansion history given by the Hubble parameter to explain that power deficiency in the low CMB multipoles. The effect of such 
modification in the Hubble parameter on other cosmological observations such as matter power spectrum,
cosmological distance modulus and galaxy cluster count is also discussed. A comparative analysis of the power 
deficiency caused by ISW effect and the primordial power spectrum has also been shown to be distinguishable with CMB 
polarization measurements at low multipoles. 

The paper is organized as follows. In section I we review the basic of ISW effect. In section II 
we present the features in the expansion history that provide a negative ISW contribution to the 
low multipoles of the CMB power spectrum. In the third section we discuss the effects of these features in the expansion history 
on other cosmological observables. In the fourth section we discuss the 
corresponding observations that
can distinguish the low power at low multipole caused by the suggested 
modification in the expansion history from 
an origin in primordial power spectrum (PPS) such low multipole
effects. The final section is devoted to discussion and the conclusions. 

\section{Understanding the ISW effect}

%The CMB power spectrum is one of the most precisely measured quantity
%in the theoretical astrophysics. There is no simple straight forward
%analytical expression for calculating the CMB power spectrum. 
The source term for computing the CMB power spectrum mainly consists of
three independent components. The Sachs Wolf effect comes from the
gravitational redshift at LSS, the Doppler term comes from the velocity
perturbation at LSS and the third part i.e. ISW term comes from the
gravitational redshift between the LSS and
the present era. An analytical expression for calculating the CMB power spectrum
\cite{Dodelson2003d,Peebles1994,Das-2010} can be written as 

\begin{equation}
C_{l}=\int_{0}^{\infty}|\Delta_{l}(k)|^{2}P(k)k^{2}dk\,,\label{eq:Cl}
\end{equation}

\noindent where, $\Delta_{l}(k)$  and $P(k)$ are the brightness fluctuation function 
and the primordial power spectrum from the inflationary scenario for wave number $k$.
$\Delta_{l}(k)$ can be written in terms of the
temperature source term, $S_{T}(k,\tau)$ and the spherical Bessel
function, $j_{l}(x)$ of order $l$ as 

\begin{equation}
\Delta_{l}(k)=\int_{0}^{\tau_{0}}S_{T}(k,\tau)\, j_{l}(k(\tau_{0}-\tau))d\tau\,,\label{eq:Brightness}
\end{equation}

\noindent where, $\tau$ is the conformal time and $\tau_{0}$ represents the conformal
time at the present epoch i.e. at redshift $z=0$.
The exact expression for the temperature source term in synchronous
gauge is given by 

\begin{equation}
S_{T}=g\left(\frac{1}{4}\delta_{g}+2\dot{\alpha}+\frac{\dot{\theta}_{b}}{k^{2}}+\frac{\Pi}{16}+\frac{3\ddot{\Pi}}{16k^{2}}\right)
+\dot{g}\left(\frac{\theta_{b}}{k^{2}}+\alpha+\frac{3\dot{\Pi}}{8k^{2}}\right)+e^{-\mu}\left(\dot{\eta}+\ddot{\alpha}\right)+\ddot{g}\frac{3\Pi}{16k^{2}}\,,
\label{eq:Source_term}
\end{equation}

\noindent where $\mu$ is the optical depth at time $\tau$. $g$ is visibility
function and is given by $g=\dot{\mu}\exp(-\mu)$. $\delta_{g}$ is
the photon density perturbation i.e. $\delta_{g}=\delta\rho_{g}/\rho_{g}$,
where $\rho_{g}$ is the density of photons. $\theta_{b}=kv_{b}$,
where $v_{b}$ represents the velocity perturbation of the baryons.
$\alpha$ is given by $\alpha=\left(\dot{h}+6\dot{\eta}\right)/k^{2}$. 
Here, $h$ and $\eta$ are metric perturbation variables 
in the $k$ space and are given by \cite{Ma1994} 
$h_{ij}(\vec{x},\tau)=\int dk^{3}e^{i\vec{k}.\vec{x}}\left\{ \hat{k}_{i}\hat{k}_{j}h(\vec{k},\tau)\right.$ 
$\left. +\left(\hat{k}_{i}\hat{k}_{j}-\frac{1}{3}\delta_{ij}\right)6\eta(\vec{k},\tau)\right\}$
and $\vec{k}=k\hat{k}$. The line element is given by $ds^{2}=a^{2}(\tau)\left\{ -d\tau^{2}+\right.$
$\left.\left(\delta_{ij}+h_{ij}\right)dx^{i}dx^{j}\right\} $,
and $a(\tau)$ is the scale factor. The indices $i$ and $j$ run from $1$ to
$3$. $\Pi$ is the anisotropic stress and in most of the cases $\Pi$
and its derivatives i.e. $\dot{\Pi}$ and $\ddot{\Pi}$ can be neglected
because they are small in comparison to other terms. In all the
expressions over-dot ($\dot{x}$) denotes the derivative with respect
to conformal time $\tau$. 

The first term in the bracket in Eq.(\ref{eq:Source_term}) can be
interpreted in terms of the fluctuations in the gravitational potential
at LSS and is referred as the Sachs-Wolfe
(SW) term. The second term known as the Doppler term, and arises due to
the velocity perturbation of the
photons at LSS. The third term provides
an integral over the perturbation variables along the line of sight
to the present era. This can be interpreted in terms of variations
in the gravitational potential along the line of sight and this is
referred as the Integrated Sachs-Wolfe (ISW) term. 

The visibility function $g$ and its derivative $\dot{g}$ peak only
at the surface of the last scattering in absence of re-ionization.
Therefore, the velocity
and the SW term are only important at LSS.
As the ISW part is not multiplied with any such visibility
function therefore it is important throughout the expansion history.
The ISW part can be broken in two parts, 1) the ISW effect close to
the surface of last scattering or the early ISW effect and 2) well after
the surface of last scattering or the late ISW effect. Therefore,
the total source term can be broken into two independent
parts, depending on the time of its creation 

\begin{equation}
S_{T}(\tau,k)=S_{T}^{Pri}(\tau,k)+S_{T}^{ISW}(\tau,k)\,,\label{eq:source_term_break}
\end{equation}

\noindent where the primordial part, i.e. $S_{T}^{Pri}(\tau,k)$ consists of
the SW, Doppler and the early ISW part and $S_{T}^{ISW}(\tau,k)$
part consist of the late time ISW part. The late time ISW term arises due to the 
presence of dark energy (DE), causing late time acceleration of the universe. 
%As the dark energy only dominates
%at the late time in the universe, it only affects the $S_{T}^{ISW}(\tau,k)$ term. 
The primordial source term is completely unaffected by DE and independent of 
late time evolution of the universe expect for reionization considered. 
%the dark energy provided there is no re-ionization. 

Using three distinct parts Eq.(\ref{eq:source_term_break}), Eq.(\ref{eq:Cl}) and Eq.(\ref{eq:Brightness}) we
%can clearly see that there are %get three independent terms in 
separate the contribution to $C_{l}$. 

\begin{equation}
C_{l}=C_{l}^{Pri}+C_{l}^{ISW}+2C_{l}^{Int}\,,\label{eq:Cl_break}
\end{equation}
%The 
\noindent where the first term, is the contribution from primordial part and is 

\begin{eqnarray}
C_{l}^{Pri} & = & \int_{0}^{\infty}\left|\Delta_{l}^{Pri}(k)\right|^2P(k)k^{2}dk\,.\label{eq:Cl_Pri}
%k^{2}dk\left\{ \int_{0}^{\tau_{0}}\left[g\left(\frac{1}{4}\delta_{g}+2\dot{\alpha}+\frac{\dot{\theta}_{b}}{k^{2}}\right)\dot{g}\left(\frac{\theta_{b}}{k^{2}}+\alpha\right)\right]\, j_{l}((\tau_{0}-\tau)k)d\tau\right.\nonumber \\
% &  & \left.+\int_{0}^{\tau_{*}}\left[e^{-\mu}\left(\dot{\eta}+\ddot{\alpha}\right)\right]\, j_{l}((\tau_{0}-\tau)k)d\tau\right\} ^{2}\,.\label{eq:Cl_Pri}
\end{eqnarray}

% The integrand being a square term is always positive. 
\noindent The second term 

\begin{equation}
C_{l}^{ISW}=\int_{0}^{\infty}\left|\Delta_{l}^{ISW}(k)\right|^2P(k)k^{2}dk\,,\label{eq:Cl_ISW}%k^{2}dk\left\{ \int_{\tau_{*}}^{\tau_{0}}\left[e^{-\mu}\left(\dot{\eta}+\ddot{\alpha}\right)\right]\, j_{l}((\tau_{0}-\tau)k)d\tau\right\} ^{2}\,,\label{eq:Cl_ISW}
\end{equation}

\noindent is the contribution from the late time ISW part. The integrand of Eq.(\ref{eq:Cl_Pri}) and Eq.(\ref{eq:Cl_ISW})
both being square terms are always positive. %and is also positive due to the similar reason. 
The third is the interference term between the Primordial and ISW source terms and given by 

\begin{figure}
\centering
\includegraphics[width=0.99\columnwidth]{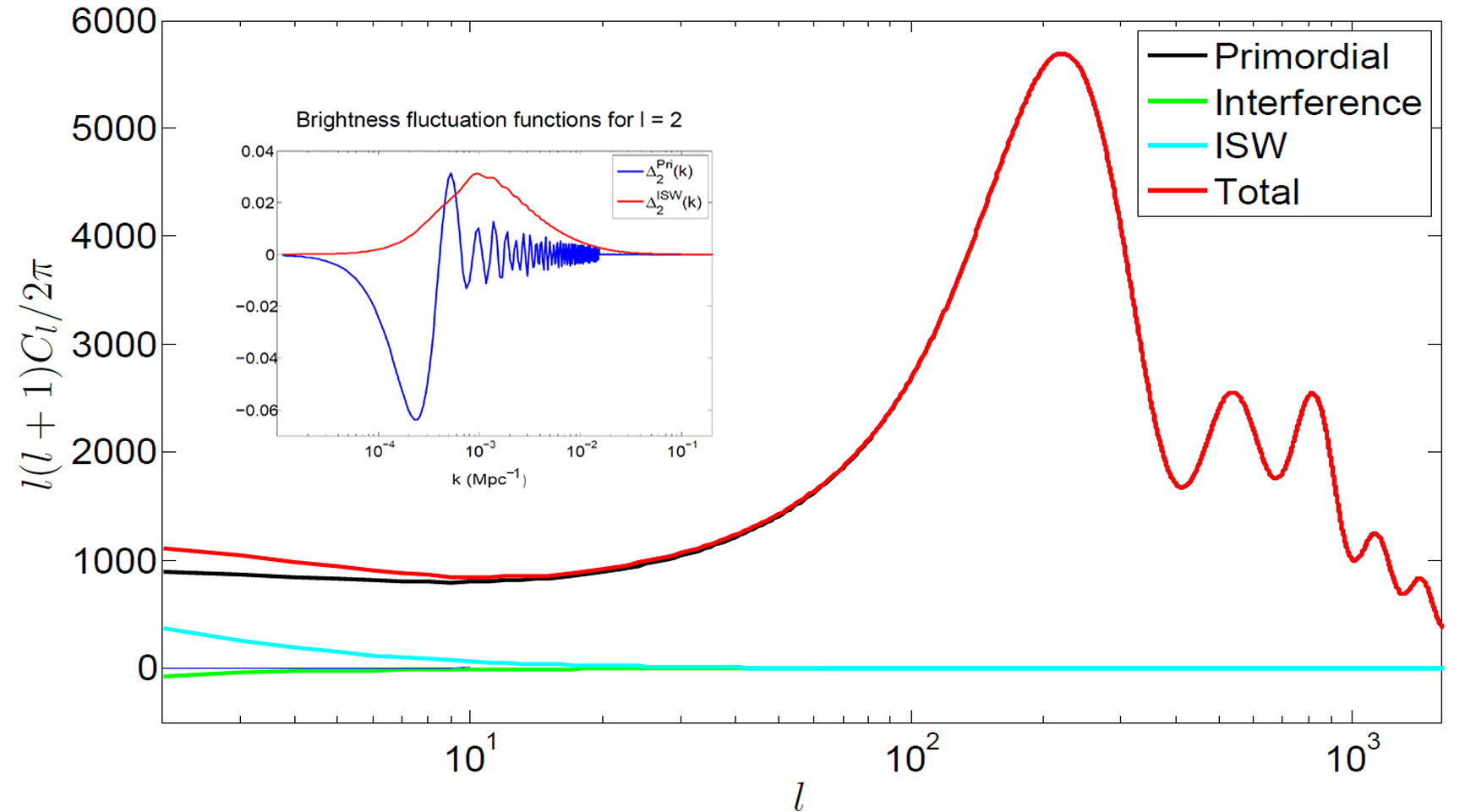}
\caption{\label{fig:Cl_LCDM}$C_l^{TT}$ for the standard $\Lambda$CDM model. 
Black plot is the $C_{l}^{Pri}$,
cyan plot is $C_{l}^{ISW}$ and the green plot is the $C_{l}^{Int}$.
Red plot is showing the total $C_{l}=C_{l}^{Pri}+C_{l}^{ISW}+2C_{l}^{Int}$.
At low multipoles the power from the primordial part is almost constant.
The increase in the power at low multipoles come from the ISW
and the interference part.}
\end{figure}

\begin{equation}
C_{l}^{Int}=\int_{0}^{\infty}\Delta_{l}^{Pri}(k)\Delta_{l}^{ISW}(k)P(k)k^{2}dk\,.\label{eq:Cl_Inte}
\end{equation}

\noindent The interference term $C_{l}^{Int}$ is an important quantity because
unlike the other two terms $C_{l}^{Int}$ can either be positive, or,
negative. %If we separate out the three terms then it can be seen that 
For $\Lambda$CDM universe, the interference term is actually negative
but is very small in magnitude in comparison to the positive $C_{l}^{ISW}$ part. 
Therefore, total contribution from the ISW term ($C_{l}^{ISW}+2C_{l}^{Int}$)
increases the power at the low $C_{l}$ multipoles with respect to the surface term. 

In Fig.~\ref{fig:Cl_LCDM} we plot all the three components for the $\Lambda CDM$ model.
% are shown independently. It can be seen that 
The Primordial part (black line) is almost flat at the low multipoles and the excess power at
the low multipoles is coming from the ISW part (cyan line). The %interface part is small because the 
first peak of %primordial brightness fluctuation (
$\Delta_{l}^{Pri}$ and the peak 
of the $\Delta_{l}^{ISW}$ are not in same phase as shown in the inset in Fig.~\ref{fig:Cl_LCDM}. 
This explains the small contribution from the interference part.
%Therefore, in the product the terms cancel 
%each other (see figure (\ref{fig:Cl_LCDM})). 
With a proper choice of 
%Now it might be possible to choose 
expansion history, the peaks of the two brightness fluctuation functions 
$\Delta^{Pri}$ and $\Delta^{ISW}$ in Eq.(\ref{eq:Cl_Inte}) can be brought in phase. In that case, the order of $\left|\Delta_{l}^{Pri}(k)\right|$
being greater than $\left|\Delta_{l}^{ISW}(k)\right|$, we shall get 
$\left|2C_{l}^{Int}\right|>\left|C_{l}^{ISW}\right|$
and their signs will be opposite resulting $C_{l}^{ISW}+2C_{l}^{Int} < 0$.
Therefore, the ISW effect, then would have a negative contribution to
the low multipole power in the CMB power spectrum.

\section{Decreasing the power at low CMB multipole}

\begin{figure}
\centering
\includegraphics[trim=0.0cm 0.1cm 1.0cm 0.0cm, clip=true,width=0.80\columnwidth]{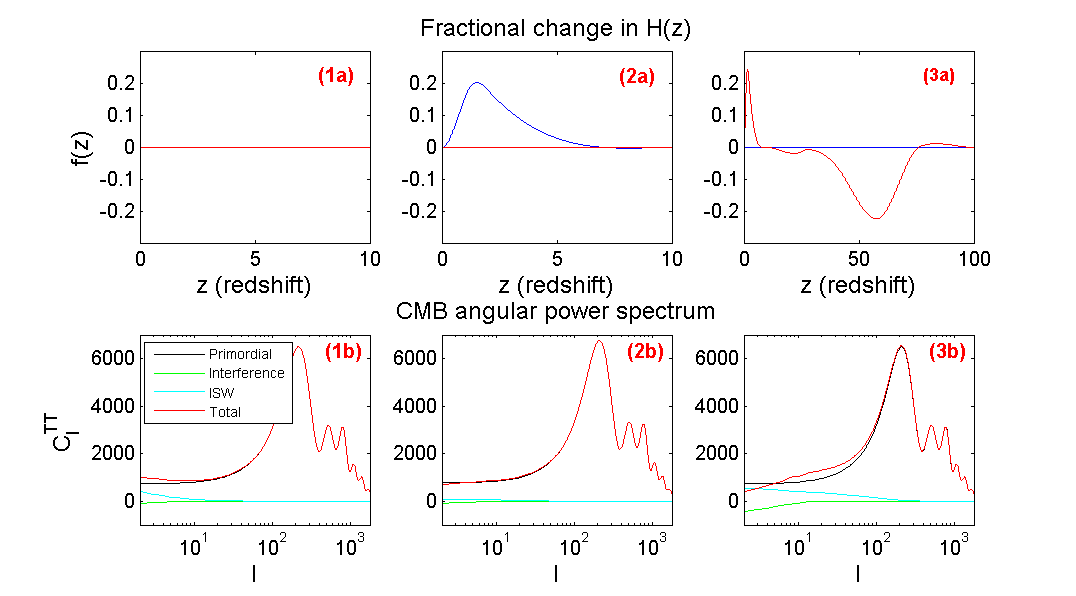}
\includegraphics[trim=1.1cm 0.0cm 2.3cm 0.0cm, clip=true,width=0.76\columnwidth]{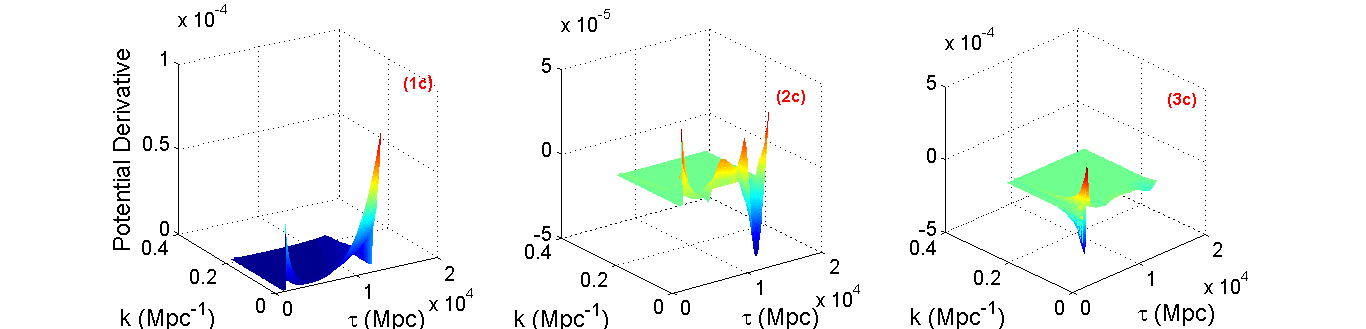}\\[0.3cm]
\includegraphics[trim=1cm .1cm 2.5cm 0.0cm, clip=true,width=0.75\columnwidth]{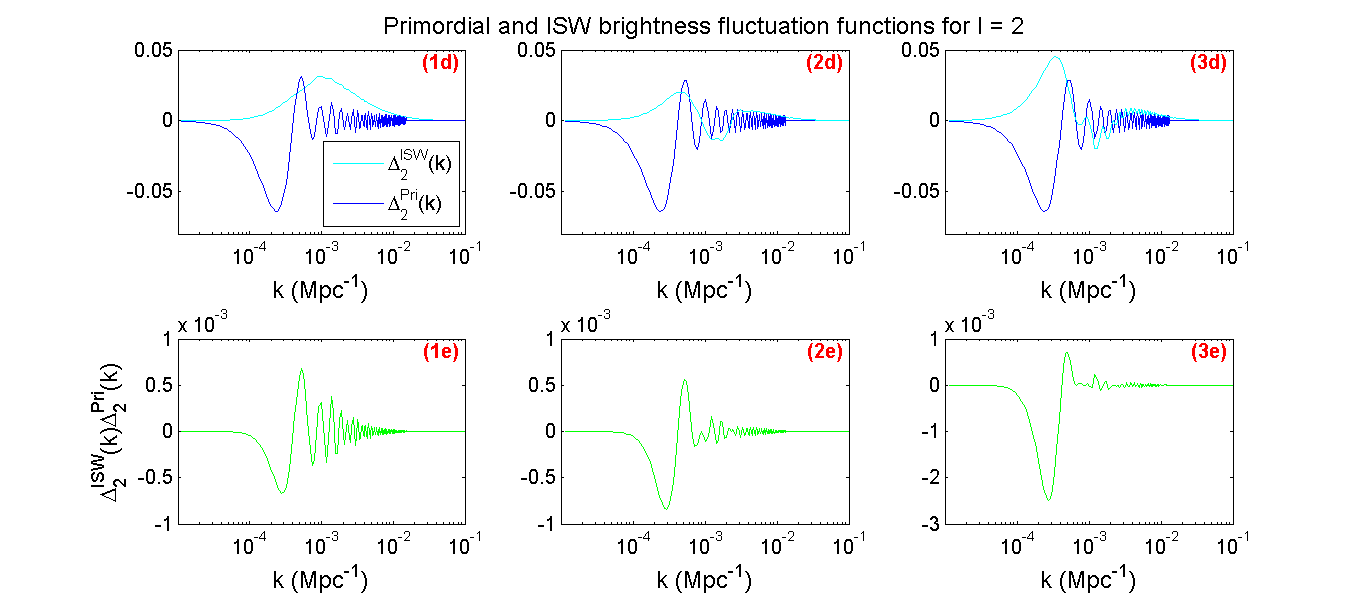}
\caption{\label{fig:LCDM_set}
{\small
Three columns of plots are shown for three different expansion histories.
The first column is for the $\Lambda$CDM model, second is for 
$\Lambda$CDM model with a bump in $f(z)$ and last column is for 
`bump + dip' model. The five rows show
$f(z)$, $C_l^{TT}$ with primordial and ISW components separated, 
ISW source term, ISW and primordial
brightness fluctuation functions for $l=2$ multipoles 
and $\Delta_2^{ISW}(k)\Delta_2^{Pri}(k)$ respectively. The 2nd and 3rd columns correspond
to the two modified expansion histories, where
ISW effect decreases low multipole power. Here we take a larger value $f(z)$ by about a factor 
of 2 than what is required to explain Planck results, to make the effects readily visible. 
}
}
\end{figure}

%According to the discussion
As discussed,
in the previous section, ISW effect can decrease the power at the low
CMB multipoles because of the interference term that can take both
positive or negative values, depending upon the expansion history of
the universe. In general, the peak of the brightness fluctuation
function of the ISW term is out of phase with that of
the primordial part, hence they cancel each
other in the integration over the wave number $k$. However, 
when the expansion history of the universe is chosen
in such a way that the peak of the ISW part is in the same
phase with that of the primordial part then the interference term
may not be very small. The interference term in that case becomes
larger in magnitude than the ISW term and with typically with a negative sign
since potentials decay during the accelerated expansion caused by dominance of any form of DE. So, 
if we add the three parts, namely primordial, ISW and the interference
together then we can obtain lower power at the low
CMB multipoles.

From Eq.(\ref{eq:Source_term}), we know that the ISW source term
is $e^{-\mu}\left(\dot{\eta}+\ddot{\alpha}\right)$. We are only interested
in the late time expansion history of the Universe. For
simplicity, we consider a case of no re-ionization only for providing 
analytical understanding where the optical
depth i.e. $\mu = 0$ after the surface of last scattering (sls). Therefore, $e^{-\mu}=1$
and hence the ISW source term in our region of interest is given by
$\left(\dot{\eta}+\ddot{\alpha}\right)$. Also $\dot{\eta}$ is given
by the expression \cite{Ma1994,Das-2010}

\begin{equation}
\dot{\eta}=\frac{1}{2k^{2}}\left[\frac{3H{}_{0}^{2}}{c^{2}}\frac{\theta_{b}}{a}+\frac{8\pi G}{c^{2}}\frac{4\sigma_{b}}{c^{3}}T_{CMB}^{4}\left(\frac{\theta_{\gamma}}{a^{2}}+\frac{7}{8}\times\left(\frac{4}{11}\right)^{4/3}N_{\nu}\frac{\theta_{\nu}}{a^{2}}\right)\right]\,,
\end{equation}

\noindent where $\theta_{b}$, $\theta_{\gamma}$ and $\theta_{\nu}$ are the
divergence of the fluid velocity and are given by $kv_{b}$, $kv_{\gamma}$
and $kv_{\nu}$, where $v_{b}$, $v_{\gamma}$ and $v_{\nu}$ are the
velocity perturbation of the baryon, photon and the neutrinos respectively
and $k$ is the wave number. $N_{\nu}$ is the effective number of neutrino
species. %All other parameters are in their standard meaning.
The baryons are decoupled from the
photons at LSS and thus $\theta_{b}$ is very small after LSS. Secondly, the contribution
from the radiation part is completely negligible due to the factor
before it. So $\dot{\eta}$ will be almost negligible after
the surface of the last scattering and the only dominating term in
the late time ISW effect will be $\ddot{\alpha}$. A mathematical expression
for $\ddot{\alpha}$ is given by \cite{Ma1994,Das-2010} 

\begin{equation}
\ddot{\alpha}=-\frac{3\dot{\sigma}}{2k^{2}}+\dot{\eta}-2\left(\frac{\dot{a}}{a}\right)\dot{\alpha}-2\frac{d}{d\tau}\left(\frac{\dot{a}}{a}\right)\alpha\,.
\label{alphadd}
\end{equation}

In the above expression, $\sigma$ is the shear term its contribution
is significantly small after LSS. 
%Similarly,
%according to the previous discussion $\dot{\eta}$ is also very small
%after surface of the last scattering. 
Therefore, the first two terms can be neglected and 
the remaining part of the Eq.(\ref{alphadd}) can be written as 

\begin{equation}
\ddot{\alpha}=-2\frac{d}{d\tau}\left(\frac{\dot{a}}{a}\alpha\right)\,.
\end{equation}

\noindent Now we using the separation of variable to break $\alpha$ apart
into a $\tau$ dependent part and a $k$ dependent part, can
write $\alpha$ as 

\begin{equation}
\alpha(k,\tau)=\alpha_{\tau}(\tau)\alpha_{k}(k)\,.
\end{equation}

\noindent The equation in the conformal time domain becomes,

\begin{equation}
\ddot{\alpha}_{\tau}=-2\frac{d}{d\tau}\left(\frac{\dot{a}}{a}\alpha_{\tau}\right)=-2\frac{d}{d\tau}\left(Ha\alpha_{\tau}\right)\,.\label{eq:alpha_t_ddot}
\end{equation}

% For the last expression we use 
\noindent where, $H=\frac{\dot{a}}{a^{2}}$ (dot represents the 
derivative wrt the conformal time). Considering
$H{}_{\Lambda}$ to be the Hubble parameter from the standard $\Lambda$CDM
model, we can define $H(a)=H_{\Lambda}(a)(1+f(a))^{\frac{1}{2}}$ where $H(a)$
is the Hubble parameter from any model and $\frac{1}{2}f(a)$ is the fractional
deviation of the Hubble parameter from the $\Lambda$CDM
model. Hence Eq.(\ref{eq:alpha_t_ddot}) becomes, 

\begin{equation}
\ddot{\alpha}_{\Lambda\tau}+\delta\ddot{\alpha}_{\tau}=-2\left(1+\frac{1}{2}f\right)\frac{d}{d\tau}\left(H_{\Lambda}a\alpha_{\tau}\right)-\left(H_{\Lambda}a\alpha_{\tau}\right)\dot{f}\,,\label{eq:alphaddot}
\end{equation}

\noindent where $\ddot{\alpha}_{\Lambda\tau}$ is the ISW source term for the
standard $\Lambda$CDM model. The equation above shows that the change
in the standard model Hubble parameter directly reflects on the ISW
source term. If $f$ is chosen in such a way that the second
term of Eq.(\ref{eq:alphaddot}) i.e. $\dot{f}$ is small
then change in $\ddot{\alpha}_{\tau}$ linearly depends on $f$. 
So, by taking the value of $f(a)$ to be positive (negative) there is a decrease (increase)
in the value of $\ddot{\alpha}_{\tau}$. 
%It means that if $f(a)$ is taken to be negative at some redshift, $\ddot{\alpha}_{\tau}$
%will decrease there, and the opposite will happen if $f(a)$ is taken
%to be positive. 
So we have a direct control over the ISW source term
by controlling $f(a)$.

To find out the form of $f(a)$ that can provide lower power at the
low multipoles we need to check the behaviour of 
%become familiar with the nature 
the ISW source terms in the standard $\Lambda$CDM model.
In Fig.~\ref{fig:LCDM_set} we plot the power spectrum from three 
different expansion histories in three different columns. The first column is for 
the standard $\Lambda$CDM model, in the second column we plot the power spectrum 
for a model with a bump in $f(z)$ and the last column is for the `bump + dip' model.
In each of the columns we show five
different quantities. The first row shows the expansion history
of the universe in terms of $f(z)$. In the second row we show  
the CMB temperature power spectrum where we separate out three different components of the 
$C_l^{TT}$. The red curve shows the
complete power spectrum. The black part shows the primordial
part, blue corresponds to ISW and green to the interference part. In the third row
we show the ISW source term i.e. $e^{-\mu}(\dot{\eta}+\ddot{\alpha})$
(derivative of the Newtonian potential in conformal gauge). In the fourth 
row we show the brightness fluctuation functions for the primordial and 
the ISW part separately for $l=2$ and in the fifth row we plot the 
$\Delta_2^{ISW}(k)\Delta_2^{Pri}(k)$ because this is the quantity which
upon integration over $k$ gives $C_2^{Int}$. 
%Hence this is one of the most important quantity.

From the first column we can see that the ISW effect increases the power at 
the low CMB multipoles. %Now to get a
%situation where the ISW part is contributing a negative power to the
%power spectrum we need to do something which can decrease the power
%of the blue curve and increase the negative power at green curve at
%low multipoles. This can be done as follows. 
In panel (1c) of Fig.~\ref{fig:LCDM_set} we plot the
quantity $\left(\dot{\eta}+\ddot{\alpha}\right)$, which in
the Newtonian gauge is actually the derivative of the Newtonian potential
with respect to the conformal time, i.e. $\left(\dot{\Psi}+\dot{\Phi}\right)$.
It shows that $\left(\dot{\eta}+\ddot{\alpha}\right)$
or rather $\ddot{\alpha}$ is increasing at the late time for the
small $k$. The brightness function for the ISW effect is just the
convolution of this $\ddot{\alpha}$ with $j_{l}\left(k\left(\tau_{0}-\tau\right)\right)$
over conformal time $\tau$. Here $j_{l}(x)$ is the spherical Bessel
function of order $l$. In general any spherical Bessel function $j_{l}(x)$
has a peak near $x=l$. Therefore, when we convolve a function with
$j_{l}\left(k\left(\tau_{0}-\tau\right)\right)$, only the value of the
function near $l=k\left(\tau_{0}-\tau\right)$ plays an important role 
%factor as most of the contribution of the function 
in the integral 
%comes from this particular part of the function. 
Now from panel (1c) 
of the Fig.~\ref{fig:LCDM_set} we know that the value of the function
$\left(\dot{\eta}+\ddot{\alpha}\right)$ is high near $\tau=\tau_{0}$,
i.e. most of the contribution of this function is located at a place where
$\left(\tau_{0}-\tau\right)$ is very small. Therefore, when we convolve
this function with $j_{l}$ for any $l$, the brightness fluctuation
function, which is a function of $k$ will have a peak near $k\sim l/(\tau_{0}-\tau)$,
i.e. at some high $k$. The $l=2$ brightness fluctuation from the
primordial and the ISW parts are shown in panel (1d) of Fig.~\ref{fig:LCDM_set}.

It shows that the peak of the  $l=2$ ISW brightness fluctuation
function is located at high $k$ (at $~10^{-3} Mpc^{-1}$) whereas the locus of the peak of the Primordial part is
at low $k$ (at $~2 \times 10^{-4} Mpc^{-1}$). Therefore, the effect of the ISW is more where the Primordial
part has an oscillating tail. Hence the product of the ISW
and Primordial brightness fluctuation, that makes up the interference term, is largely %we are making the entire product
oscillatory and on integration over $k$ leads to a very small
contribution only from low $k$. The rest of the part, i.e.
high $k$ part after integration is almost zero. The product
of the $l=2$ brightness fluctuation functions i.e $\Delta_2^{Pri}(k)\Delta_2^{ISW}(k)$ is shown in panel (1e).
From the plots it is apparent that to lower the power at the low multipoles it is important
to shift the peak of the ISW brightness
fluctuation function towards the low $k$ such that $\Delta_2^{Pri}(k)\Delta_2^{ISW}(k)$ become less oscillating. 

The peak of the brightness
fluctuation function of order $l$, roughly located near $k\sim l/(\tau_{0}-\tau_{*})$,
where $\tau_{*}$ is the conformal time of some effective position of the peak in
the ISW source term in the conformal time domain. So to shift the peak of the ISW brightness fluctuation
function to low $k$ we need an effectively large $\left(\tau_{0}-\tau_{*}\right)$.
This implies that we need to reduce the power in the ISW source term
at low $\left(\tau_{0}-\tau\right)$ and increase the power at high
$\left(\tau_{0}-\tau\right)$, %which implies %i.e. we need 
to shift the effective peak of 
ISW source term towards low conformal time. The ISW source term 
is mostly given by $\ddot{\alpha}$. %Previous discussions
%show us the changes required in the expansion history of the universe
%which can provide some special kind of change in the ISW source term.
So, by putting some bump and dip like features in the expansion history, 
the ISW source term and the power from the ISW term in CMB power spectrum 
can be controlled. The exact results for the two illustrative modification to $H(z)$ 
shown in Fig.~\ref{fig:LCDM_set} clearly bears out the 
expectations from our approximate analytical
calculations.

The middle column of Fig.~\ref{fig:LCDM_set} shows %what happens when we put 
effect of a bump like feature in the $f(z)$
at some low redshift. Panel (2a) of Fig.~\ref{fig:LCDM_set} shows the shape
of $f(z)$. According to our previous discussion a bump like
feature in $f(z)$ will decrease $\ddot{\alpha}$ near $\tau \approx \tau_{0}$ 
and is given in panel (2c) of Fig.~\ref{fig:LCDM_set}.
In the fourth row it can be seen that the brightness
fluctuation function is decreasing at high $k$, which is a direct
effect of decrease in $\ddot{\alpha}$ and is expected
from our analytic arguments. %the previous analysis. 
This decrease of ISW source term supress
the power of the low multipoles of $C_{l}^{ISW}$. 
The interference term is not large due to cancellation 
of ISW and primordial part at high $k$ regime.
%of much importance
%because the interference term doesn't get much contribution from the
%high $k$ part of the brightness function as it gets canceled
%when being multiplied with the primordial part.
%As the 
Due to decrease in $C_{l}^{ISW}$ at low multipoles, the combined effect
of $C_{l}^{ISW}$ and $C_{l}^{Int}$ provides a negative
contribution at low CMB multipoles. 
In Fig.~\ref{fig:LCDM_set} we plot these combined effect and its implication on total $C_l$ .
However, in this case the distance to LSS is affected leading to shifts in the $C_l$ spectra 
unless $H_0$ is chosen much smaller than the measured value.

%The effect is clearly visible
%from $C_l^{TT}$ plot in the middle column of figure(\ref{fig:LCDM_set}). 

To reduce the power further at the low multipoles
of $C_{l}{}^{TT}$ we can add a dip like feature in $f(z)$ at some
middle redshift ($z \approx 20-50$), which also increases the
power at early time in the ISW source function. The shape of the `bump + dip' model of $f(z)$ is
shown in panel (3a) of Fig.~\ref{fig:LCDM_set}. This 
modification to the Hubble parameter for standard $\Lambda$CDM 
provides high value of the source function ($\ddot{\alpha}$) at
high redshift (panel 3c of Fig.~\ref{fig:LCDM_set}). 
As argued above, this particular
effect increases the value of the CMB brightness fluctuation
function at the low $k$ (panel 3d Fig.~\ref{fig:LCDM_set}). put
Therefore, the peak location of the ISW
source term and the Primordial source term are now arranged to be even closer in phase at
the low $k$. Here, the interference term due to the product 
of the ISW and Primordial, i.e.
the cross term becomes even more negative at the low multipoles 
(see panel 3c of Fig.~\ref{fig:LCDM_set}). Consequently, the interference
part results in the power at low multipoles to decrease further in this particular
case. Its clear effect on $C_{l}$ can be seen in the last row of Fig.~\ref{fig:LCDM_set}.
%So as detailed in the previous section we need 
Another important point is that by putting  bump in $f(z)$ 
at low redshift and a compensating dip in $f(z)$ at medium redshift 
the total distance to LSS can be restored. So, by choosing a bump and
the compensating dip at proper redshifts the ISW effect can provide a low
power at the low CMB multipoles while preserving the $C_l$ spectra at high $l$
for the measured value of $H_0$.

 Here one should note that
the width $\Delta z$ of the bump is small at the low $z$ where as its large
at high $z$. This is due to the fact that the conformal time scales
with $z$ according to the formula $\Delta\tau=\frac{\Delta z}{z+1}$.
So the when $z$ is large, same $\Delta z$ corresponds to a smaller
change in the conformal time domain (i.e. $\Delta\tau$). To convert
the source function to the brightness fluctuation function the integration
is over the conformal time domain, i.e. $\tau$. %So for getting same
In order to preserve the distance to the last scattering surface, 
where we need a narrow bump at the
low redshift, we need to put a wider dip at high redshift. 
This explains the nature of the bump and the dip in the $f(z)$
in last column of Fig.~\ref{fig:LCDM_set}. 

%The second point here is %to see
%that dip at high redshift also significantly affects the higher multipoles
%in $C_{l}$. So it is important to keep the width of the dip ($\Delta\tau$)
%at high redshift significantly small. That will keep the 
%higher multipoles of $C_l^{TT}$ intact and only change the $C_{l}$
%low multipoles. 

\begin{figure}
\includegraphics[trim=0.6cm 8.5cm 1.2cm 9.2cm, clip=true, width=1.0\linewidth]{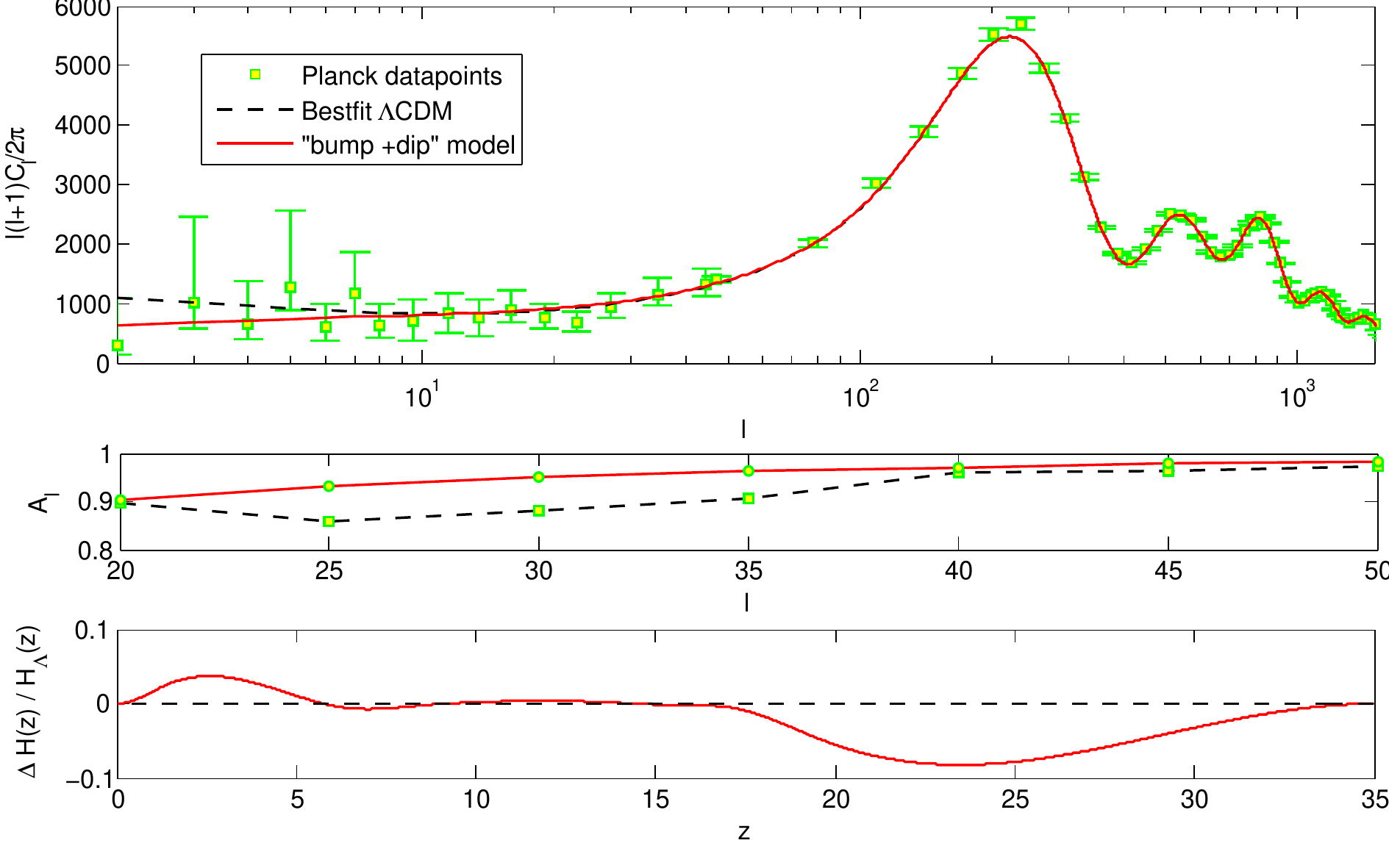}
\caption{\label{planckps}
Power spectra from the `bump + dip' model 
is plotted along with the best fit $\Lambda$CDM model
against the Planck observed data points. Planck low-l values are plotted at 2, 3, 
4, 5, 6, 7, 8, 9.5, 11.5, 13.5, 16, 19, 22.5, 27, 34.5, and 44.5 as that can provide
a better idea about the power deficiency at the low multipoles. 
The middle column shows the $A_l$, the best fit multiplication factor to match $A_lC_{l'}^{\Lambda CDM}$ 
with $C_{l'}^{Planck}$ or $C_{l'}^{model}$ between $l'=2$ to $l'=l$. In the lower plot 
we show expansion history in terms of $\frac{H(z)}{H_{\Lambda}(z)}-1$ for which the $C_l$'s are shown on the top plot.
We can see that 5-9\% deviation of $H(z)$ from standard $\Lambda$CDM model can bring the low multipole powers
significantly down. The shape of $f(z)$ is chosen just for demonstration purpose and not obtained by any parameter estimation method.  
}
\end{figure}

In Fig~\ref{planckps} we show the data points for Planck $C_l^{TT}$ in yellow boxes and 
the best fit $\Lambda$CDM model in black dotted line. Planck low-$l$ data points are plotted at 2, 3, 
4, 5, 6, 7, 8, 9.5, 11.5, 13.5, 16, 19, 22.5, 27, 34.5, and 44.5 to provide
a clearer visualization of the power deficiency at the low multipoles. It can be seen that at the low 
multipoles the observed power is lesser than the best fit $\Lambda$CDM model. 
In the second plot 
we plot the quantity $A_{l'}$, where $A_{l'}$ is the best fit 
multiplication constant $q$ such that 
$\chi^2 = \sum_{l=2}^{l'}(qC_l^{\Lambda CDM} - C_l)^2$ is minimum.
%matches best with $C_l^{Planck}$ or $C_l^{Model}$ for $2\le l \le l'$.  
The black dotted line shows the $A_l$ for Planck observed values 
%It can be seen that there is a 
clearly presenting the power deficiency at the low multipoles. If $f(z)$ is modified by putting a   
bump at low redshifts and a dip at high redshifts then the power decreases at the low multipoles and 
theoretical predictions are much closer to the observed data points from Planck. The $C_l$ and $A_{l'}$ for this 
`bump + dip' model are shown in red curve.
% shows the integrated power deficiency for a model with a bump in $f(z)$. 
A likelihood estimation
using Planck likelihood code \cite{plc} (Only for the $C_l^{TT}$) shows an improvement  
in $\Delta\chi^2 = -0.92$ for this particular illustrative model. Since the cosmic variance 
is large at the low l, the improvement in $\chi^2$ is not expected to be very large, but certainly
the illustrative $f(z)$ studied here points the way to a possible resolution of the 
observed power deficiency. Here the shape $f(z)$ 
is chosen just as a proof of concept using a GUI interface in CMBAns and not
a parameterized search of most optimal $f(z)$. 
%by any parameter estimation method. Therefore, it may not be  
The best fit $f(z)$ from such future exercise may yield further suppression of low multipole power. 
%Putting a dip in the low redshift will farther decrease the 
%power at very low multipole as discussed previously. However that will increase the power 
%little at high multipoles as can be seen from the figure (\ref{fig:LCDM_set}). 
%The plot shows that t
The purpose of this work is to point out the possibility that the low power observed by Planck
satellite may be explained by modification in the expansion 
history of the universe, thus providing a hint of richness to the nature of  
dark energy\cite{Das2012,Shafieloo2010,Shafieloo2010b}.% model discussed 
%by different authors \cite{Das2012,Shafieloo2010,Shafieloo2010b}. 

\section{Effect on other cosmological observables}   

\begin{figure}
\centering
\includegraphics[trim=0.6cm 8cm 1cm 9.5cm, clip=true, width=1.0\linewidth]{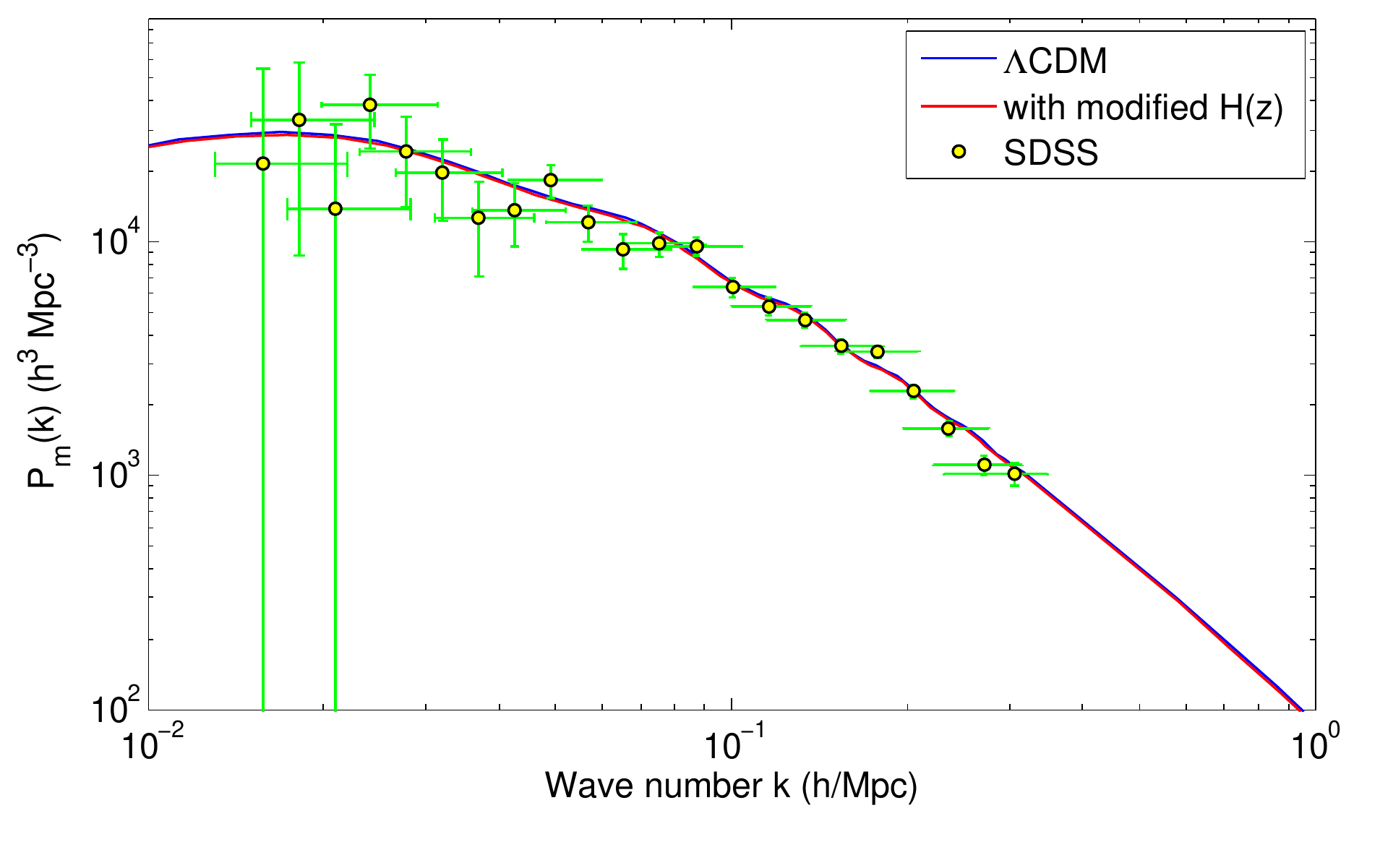}
\caption{\label{fig:mpower} The matter power spectrum for the Planck best fit $\Lambda$CDM
model and the model with a modification in $H(z)$ are shown. 
From the plots it can be seen that $P_m(k)$'s from
both the models almost overlap over each other. Therefore, matter power spectrum 
can not distinguish this model from $\Lambda$CDM model.}
\end{figure}

\subsection{Effect on matter power spectrum}
Matter power spectrum, $P_m(k)$, is an important cosmological observable for inferring the 
correct cosmological model and any %in cosmology and any physical
cosmological model should satisfy the observed Matter power spectrum ($P_m(k)$).
%  theory should satisfactorily explain this. 
In Fig.~\ref{fig:mpower} we compare the matter power spectrum 
for $\Lambda$CDM model in blue and the a model with a `bump + dip' model 
as shown in Fig.~\ref{planckps}, in red with the  
observed data points from SDSS . $P_m(k)$ from our model is consistent  
%both the models shows a very good agreement with the 
with the observed data. 
The model with modification in $f(z)$ is slightly below the $\Lambda$CDM model. Improvement 
in observation method in future can lead to a measurable change in $P_m(k)$.
%but the departure is almost insignificant. 
At this point it should be noted that, $P_m(k)$ depends of the amount of 
matter $\Omega_m = \Omega_b + \Omega_c$. Therefore, if $\Omega_{b}$ and $\Omega_{c}$ are kept 
constant then the change in the matter power spectrum will be very small. However, if such a 
shape of $f(z)$ is chosen where we need to change $H_0$ to keep the distance to the last 
scattering surface constant, such as a model with only one bump or only a dip in $f(z)$, 
then to explain $C_l^{TT}$ peaks properly we need to keep $\Omega_{b}h^2$ and $\Omega_{c}h^2$ 
constant. In that case the matter power spectrum will strongly deviate from the standard model 
$P_m(k)$ and hence the matter power spectrum can be used to distinguish the change in Hubble parameter. 

%can cause similar change in the low CMB 
%multipoles.  Though, in that case the ratio of the second 
%and third peak in $C_l^{TT}$ will change and then this effect can be distinguished from 
%other effects by $C_l^{TT}$.
 
\begin{figure}
\centering
\includegraphics[trim=2.0cm 9cm 3cm 9.5cm, clip=true, width=1.0\linewidth]{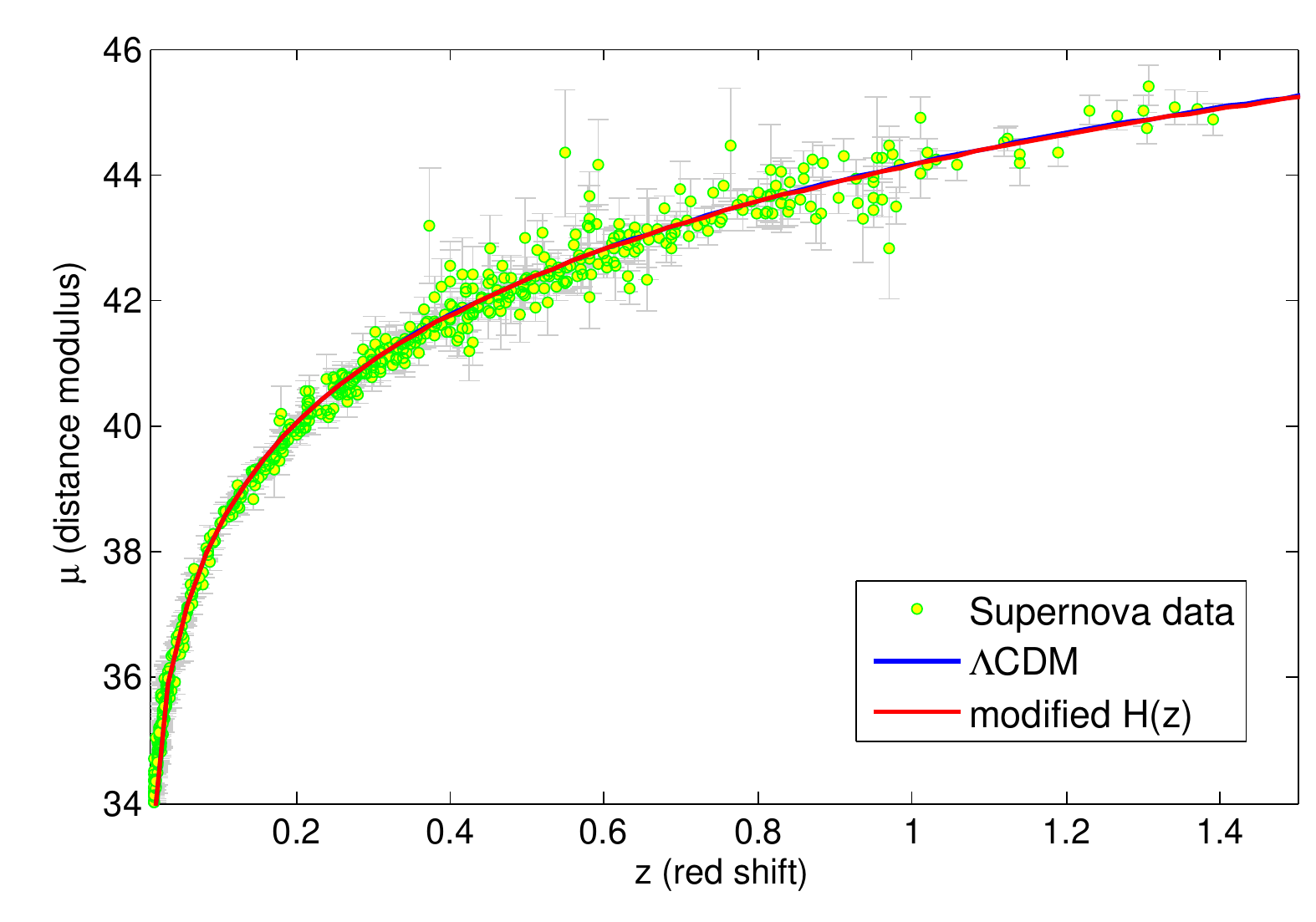}
\caption{\label{fig:Sn1} The figure shows the distance modulus for the $\Lambda$CDM 
Hubble parameter and for the model with a modification in $H(z)$. 
From the plots it can be seen that $\mu$ from both the models fall on top of each other. 
Therefore the observational dataset can't distinguish between two distance moduli.}
\end{figure}

\begin{figure}
\includegraphics[trim=2.5cm 9cm 3cm 9.5cm, clip=true, width=1.0\linewidth]{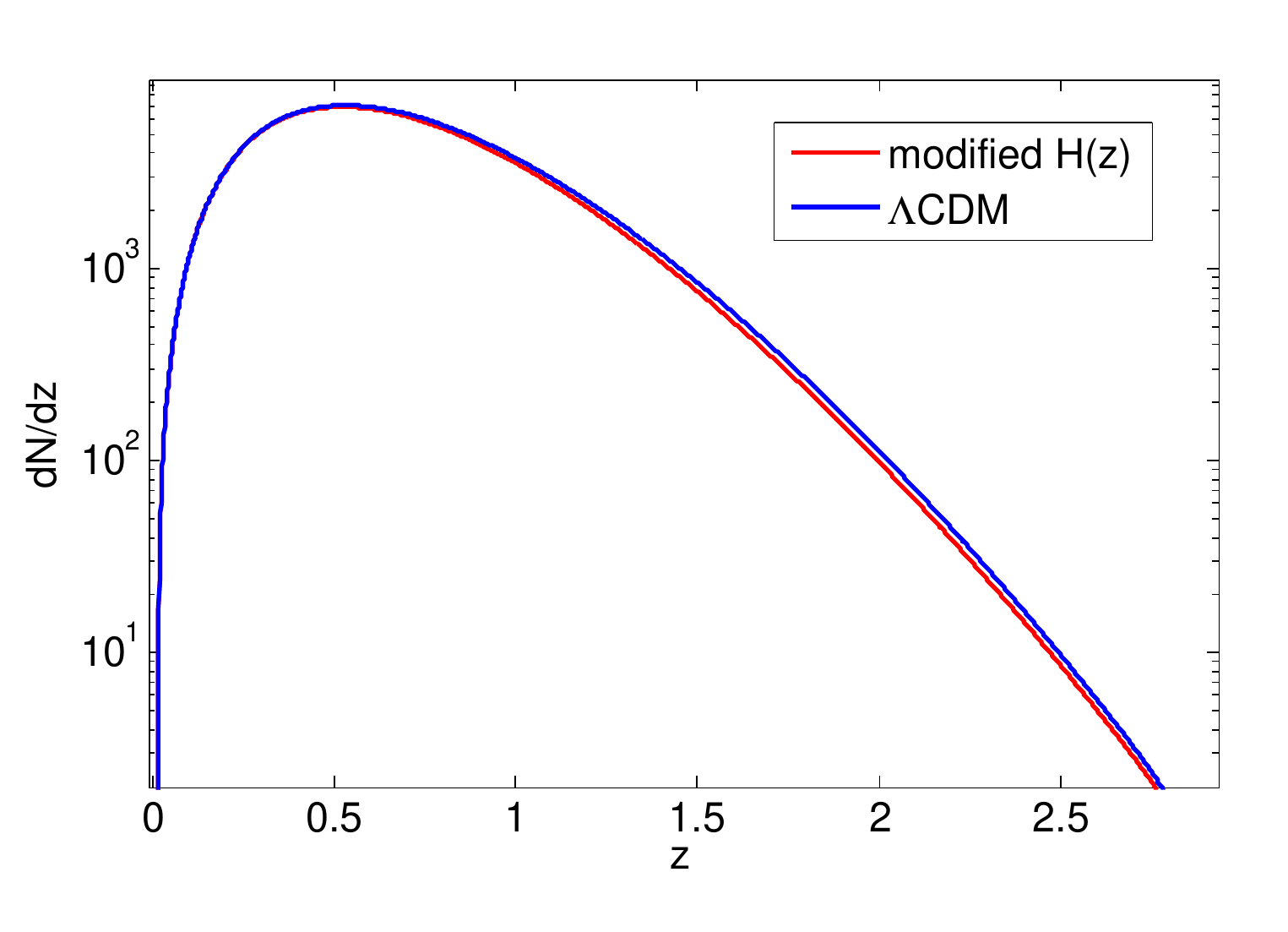}
\caption{\label{fig:dNdz} Variation of galaxy cluster number count with 
respect to redshift for the $\Lambda$CDM
model and the model with a modification in $H(z)$ are shown. 
From the plots it can be seen that $\frac{dN}{dz}$ for the modified Hubble parameter 
model is slightly lower than the $\Lambda$CDM model at high redshift. However, the difference is 
not significant enough to be detected by any observational data.}

\end{figure}

\subsection{Effect on cosmological distance modulus}
The cosmological distance modulus is another important observable in astronomy which measures 
the difference between the apparent magnitude and the absolute magnitude of an object and gives 
a measurement of the distance in astronomy. The distance modulus \cite{Shafieloo2013,Aluri2013} is defined by

 \begin{equation}
\mu=5\log_{10}d_L + 25\,,
\end{equation}

\noindent where $d_L$ is the luminosity distance of an object. Sn1a type of supernova are used 
for cosmological distance measurements and give a strong constrain on Hubble parameter and
other cosmological parameters. As we modify the Hubble parameter for lowering the ISW effect, it is important 
to estimate its effect on the cosmological distance modulus.  
%check if it introduces some observable changes on the cosmological distance modulus which can 
%distinguish the particular model from other cosmological models using present observational data. 
In Fig.~\ref{fig:Sn1} we plot the $\mu$ parameter both from $\Lambda$CDM model in blue and from 
the modified $H(z)$ as shown in Fig.~\ref{planckps} in red and its comparison with the observation
from supernova \cite{supernova1,supernova2}. 
The slight change in $\mu$ due to the modified H(z) is indistinguishable with present observation.
%From the figure it can be seen that 
%the plots are hardly distinguishable from one another. The observed data points from supernovae 
%are also plotted over the theoretical curves. 
Supernova data points %can actually 
probes the expansion history of the universe up to a very low redshift. As the present time Hubble parameter  
is kept constant therefore cosmological distance modulus unable to probe this change in the expansion
history.

\subsection{Effect on Galaxy cluster number count}

Another important cosmological observation is the evolution in the galaxy cluster number count. 
The variation of comoving number of clusters,
whose mass $M$ is greater than a fiducial mass $M_{0}$, is given
by

\begin{equation}
\frac{d\mathcal{N}}{dz}=\frac{dV(z)}{dz}N(M>M_{0},z)
\end{equation}

\noindent where $V(z)$ is the comoving volume at redshift $z$ and $N(M>M_{0},z)$
is the mass function\cite{Campanelli2012,Majumdar2004}. The comoving volume 
$V(z)$ directly depends on the luminosity distance, Hubble parameter and the 
redshift, and is given by 
\begin{equation}
V(z) = 4\pi\int_{0}^{z}dz'\frac{d_L^2(z')}{(1+z')^2H(z')}\,.
\end{equation}

As the luminosity distance $d_{L}$ does not change much, the change in the
comoving volume is not very significant. However, $N(M>M_{0},z)$
depends on the growth factor and it is given by 

\begin{equation}
N(M>M_{0},z) = \int_{M_0}^{\infty}dM n(M,z)\,.
\end{equation}

\noindent Here $n(M,z)$ is the comoving number density at redshift $z$ of clusters with masses in the 
interval $[M,M+dM]$ and it is given by \cite{Campanelli2012,Majumdar2004}

\begin{equation}
n(M,z) = \frac{2\rho_{m}}{M}\nu f(\nu)\frac{d\nu}{dM} \,,
\end{equation}

\noindent where 

\begin{equation}
\nu = \frac{\delta_c}{\sigma}\,.
\end{equation}

\noindent $\delta_c$ is the critical density constant\cite{Campanelli2012}. $\sigma$ is the amplitude of the rms density fluctuation
in  sphere of comoving radius $R$ and can be mathematically written as 

\begin{equation}
\sigma^2(R,z) = \frac{1}{2\pi^2}\int_{0}^{\infty}dk k^2 Ak^{n} T^2(k)D^2(z)W^2(kR)\,.
\end{equation}

\noindent Here $n$ is the spectral index, $D(k)$ is the growth factor, $T(k)$ is the transfer function, $W(x)$ 
is the top hat window function. 

As $N(M>M_{0},z)$ depends on the growth factor, there are some 
difference in this parameter for $\Lambda$CDM and for the modified $H(z)$ model. 
We plot the quantity $\frac{d\mathcal{N}}{dz}$
in the Fig.~\ref{fig:dNdz}. The deviation in $\frac{d\mathcal{N}}{dz}$
due to change in H(z) is dominant at higher redshift. 
%However, with present observation is consistent with both these models. 
%Improvement in future experiments may lead to observational 
%evidence for any of the particular model.
% It can be seen from the figure that the quantity deviates only
%slightly from the standard $\Lambda$CDM model at high redshift.
Although measurements of $\frac{d\mathcal{N}}{dz}$ from cluster survey are improving 
observationally this marginal variation appears to be very difficult
to discern in the near future using this cosmological probe from $\Lambda$CDM due to
modified $H(z)$.

%\section{Distinguishing ISW effect from other low multipole effects}

\section{Distinguishing low multipole power deficit generated by 
Primordial Power Spectrum (PPS) and ISW}
The effect of ISW is only limited to the CMB temperature power
spectrum and not on polarization power spectrum. The 
$E$ mode polarization source term can be written as 

\begin{equation}
S_{E}(k,\tau)=\frac{3}{16}\frac{g(\tau)\Pi(k,\tau)}{k^{2}(\tau_{0}-\tau)^{2}}\,.
\end{equation}

\noindent Since this expression does not contain any potential dependent term,
it does not depend on the expansion history, provided the distance of the
last scattering surface remains same. However, if an some expansion
history is chosen which changes the distance of the LSS from the present era
then $E$ mode source term changes due to $(\tau_{0}-\tau)$
factor in its denominator, which results into shift of $E$-mode 
polarization power spectrum %will also be shifted along the 
towards higher or lower $l$ depending on the distance of the LSS
from the present era. %However, this is not a direct effect of ISW.

In contrast if the power deficit at low multipoles arises due to features in the 
% In general when we change the expansion history of the universe,
%the distance of the last scattering surface from the present time
%may also change depending on the expansion history of the universe.
%If the change is significant then the power will shift towards low
%or high $k$ depending on the situation. Therefore, the entire power
%spectrum will shift parallel towards low or high $l$. This
%effect can be seen in all the 4 power spectra i.e. $C_{l}^{TT}$,
%$C_{l}^{EE}$, $C_{l}^{TE}$ and $C_{l}^{BB}$. However, if the change
%in the Hubble parameter don't affect the distance of the last scattering
%surface then the effect of the ISW will only be limited to $C_{l}^{TT}$
%and $C_{l}^{TE}$ part.
%There are other effects such as 
primordial power spectrum (PPS) from 
inflation
% background can also reduce the powers at the low CMB multipoles. 
%But change in PPS changes the 
power deficit, it will affect the low multipoles of the polarization power spectra well. %all the 
%(temperature and polarization) 
Hence reliable CMB polarization spectra measures a low multipole 
will be key to power spectra establishing a possible link of low multipole
power deficit to modified expansion history.

%Therefore, this effect can be distinguished
%from the other effects like the primordial power spectrum which also
%causes the change of power at the low multipoles. The change in the
%primordial power spectrum may also change power at the low CMB multipole.
% But if the low multipole
%effect is caused by the ISW effect then that will possibly change
%the power-spectrum which are related to the temperature, as only the
%temperature source term is getting changed by the ISW effect. The
%polarization power spectrum will not get changed. So by compering
%with the observational data it can be told that which effect is causing
%the low power at low multipole effects. 

%Here we have shown the plots CMB power spectra due to the change in
%the primordial power spectra. 

\begin{figure}
\centering
\includegraphics[trim=2.4cm 1cm 3cm .1cm, clip=true, width=1.0\columnwidth]{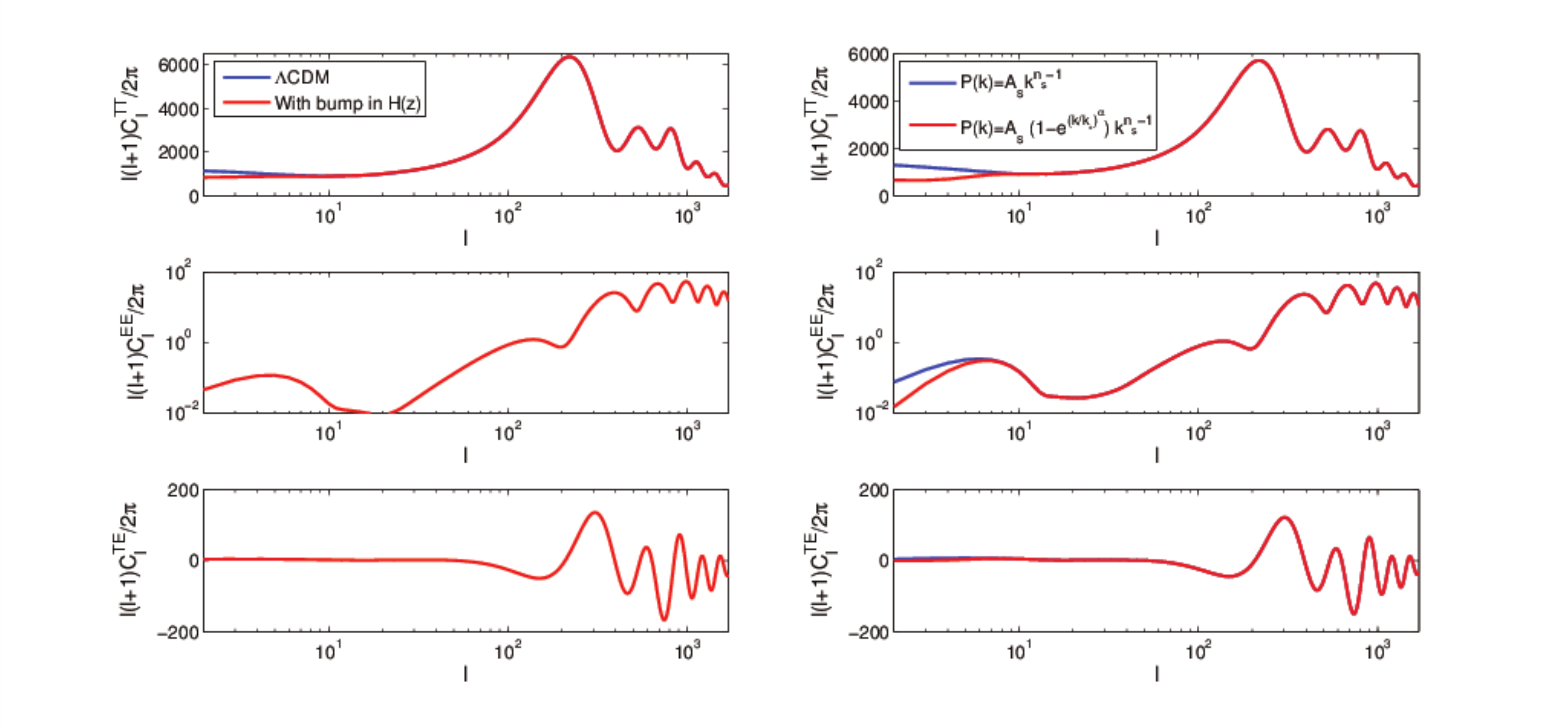}
\caption{\label{fig:Comperison}
The first column shows the change in the low multipoles due to change in the 
expansion history of the universe. The $C_l^{TT}$ and slightly $C_l^{TE}$ get 
affected due to the ISW effect, but $C_l^{EE}$ remains unchanged. In the second 
column we have shown change in the low multipole power due to the change in the 
primordial power spectrum. The effect of change in PPS is clearly visible at the low multipoles 
of both temperature and polarization power spectra.
}
\end{figure}

In Fig.~\ref{fig:Comperison} we compare the temperature and the 
polarization power-spectra for modified $H(z)$ and for modified PPS. For 
modifying $H(z)$ here we take the shape of the deviation 
as shown in Fig.~\ref{planckps}. The $\Lambda$CDM 
model power spectra are shown in the blue color and the spectra for the modified 
$H(z)$ are shown in red color. For both the cases we have used the parameter set as
$\Omega_{b}h^2 = 0.022$, $\Omega_{c}h^2 = 0.120$, $n_{s} = 0.96$, $H = 67.11$ km/s/Mpc, $\tau = 0.09$.
 Here we can see that only 
temperature and the cross power spectrum 
gets affected when ISW linked power deficit is considered.
%is getting effected due to the ISW. 
In the second column we show the change in the $C_l$ due to PPS. We take 
a modified form of PPS as 
$P(k)=A_s(1-e^{(k/k_{*})^{\alpha}})k^{n_s-1}$ \cite{Contaldi2003,Mortonson2009},
where we take $\alpha = 3.35$ and $k_{*}=7 \times 10^{-4}$ Mpc$^{-1}$.
The plots show that the change in PPS is visible in all the CMB power spectra. 
Therefore, the low power at low multipoles
caused by PPS and the ISW effect can be distinguished by combining TT and polarization spectra.% this effect. 
However, due to larger error-bars at low $l$, it may turn out to difficult 
to distinguish between effect from ISW and PPS unless the PPS power suppression is very strong.
%as the error bars are high for polarization power spectra, this may be difficult to distinguish
%between these two effects using present data. %polarization power spectra. 

\section{Discussion and conclusion}

The analytical calculations in this paper show that the ISW effect
can decrease the power at the low CMB multipoles, i.e. can provide
a negative contribution to the power at low multipoles which is not
a very well known fact. Plank data shows
that the observed power at the low CMB multipoles, particularly up
to $l=30$ are lower than the theoretical predictions of the best-fit $\Lambda$CDM
model. In this paper we demonstrate that by modifying the expansion 
history of the universe. As a proof of concept demonstration we show that
 by putting a bump a low redshift and a dip at 
high redshift in $H(z)$ as shown in Fig.~\ref{planckps}, power deficiency at low multipole
can be obtained. %This particular feature in the expansion history of the universe 
%does not introduce any observable effect in other 
We also show distinguishing this illustrative model of 
modified $H(z)$ from $\Lambda$CDM H(z) is currently well beyond the scope of other  
cosmological observables such as matter power spectrum, 
cosmological distance modulus or galaxy cluster number count. 
%Therefore, it is difficult to distinguish from other effects using different 
%cosmological dataset. 
The ISW effect does not affect the polarization power spectrum
and hence CMB polarization spectra at low multipole can be 
in principle used to distinguish this particular effect from  
power deficit originating features in the Primordial power spectrum. 

%other cosmological effects such that the change in PPS using polarization data. 
%Though, this ISW effect may be theoratically distinguished from other effects causing low 
%power at low CMB multipoles using the CMB polarization power spectra, 
%But as the the error bars in the polarization power spectra are high, in practical 
%its very much indistinguishable from PPS changes.

%The change in the matter power spectra data is almost insignifant due to this particular effect. 
%Also if the expansion history of the universe is changed without keeping the distance to LSS fixed
%then the spectra will shift towards low to high $l$. Therefore, to get a reasonable 
%power spectra along with some change in the expansion history if we change $H_0$ then  
%if $\Omega_bh^2$, $\Omega_ch^2$ are kept fixed then the matter power 
%spectrum will shift and hence the ISW effect can be distinguished using 
%matter power spectrum. However, while changing $H_0$ if we keep $\Omega_b$,
%$\Omega_c$ fixed then that will change the ratio of 2nd and 3rd peak of $C_l$s.
%Hence that will be a distinguishing factor But if we change the expansion history 
%of the universe using a combination of bump and dip in $f(z)$ such that $H_0$ remains fixed
%then the change in matter power spectrum will also become very small. 

\acknowledgments{We would like to thank Sumanta Chakraborty for his participation in some
explorations as a part of summer project, Pavan Aluri for proving the supernova dataset and 
Suvodip Mukherjee for a careful reading of the manuscript.
S.D. acknowledge the Council of Scientific and Industrial Research (CSIR), India for financial support 
through Senior Research fellowships. Computations were carried out at the HPC facilities at IUCAA. 
T.S. acknowledges Swarnajayanti fellowship grant of DST India.}

\end{document}